\documentclass{article}
\usepackage[a4paper, total={6.75in, 9in}]{geometry}
\usepackage{graphicx}
\usepackage{hyperref}
\usepackage[capitalise]{cleveref-usedon}
\usepackage{tikz}
\tikzset{>=latex}
\usepackage{xspace}
\usepackage{CJKutf8}
\newcommand{\neun}{\begin{CJK}{UTF8}{mj}{ㄴ}\end{CJK}\xspace}
\usepackage[
backend=biber,
style=numeric-comp,
sorting=none
]{biblatex}
\title{A Lower Bound on the Number of Fundamental Constants}
\author{William Luke Matthewson\footnote{Korea Astronomy and Space Science Institute, 776, Daedeokdae-ro, Yuseong-gu, Daejeon 34055, Republic of Korea,\\
${ }$\hspace{15pt}Email: \href{willmatt4th@kasi.re.kr}{willmatt4th@kasi.re.kr}}}
\date{}

\begin{document}
\maketitle
\vspace{-15pt}

\vspace{-5pt}

\abstract{
 We describe here, for the first time, a lower bound on the total number of fundamental constants required for a mathematical description of our physical universe to be complete. The answer is shown to be one. The formal arithmetized meta-mathematical proof of this is left to the reader.
}

\begin{center}
{
${ }$\\
{`}\textsc{ACHILLES}:\quad{This fellow must be playing the fool}.{'}\\
}${ }$
- from \textit{G\"odel, Escher, Bach} (1999) by Douglas R. Hofstadter \cite{Hof}.
\end{center}

\section{Some Fundamentals}

The issue of fundamental constants has been widely discussed in the literature. Given that the purpose of scientific inquiry is to pursue a complete understanding of the nature of reality, by means of some mathematical description, it is reasonable to ask whether there will be some necessary fundamental axioms that must be assumed, the precise details of which are determined simply by the arbitrary happenstance\footnote{To quote Lev Okun, `The word
``arbitrary'' here means that a given parameter has acquired its actual value in the process of cosmological evolution of the early universe. It could have with some probability other values as well.'~\cite{okun1991}} of our universe, in order to achieve this description.

In order to avoid getting caught up in arguments of a semantic nature, we should clarify here what is meant by a fundamental constant. The general consensus in the literature seems to be that in order to be considered a fundamental constant, a quantity should satisfy various conditions \cite{1983RSPTA.310..249W,okun1991,Duff_2002,Matsas_2024}.
Firstly, a fundamental constant is a quantity that cannot be derived from other, more fundamental, constants and which is necessary for a complete description of reality~\cite{1983RSPTA.310..249W}.
On the surface, true fundamental constants are also obliged to be dimensionful. In \cite{Duff_2002} the distinction is made between so-called `fundamental parameters' (e.g. the fine-structure constant, $\alpha = e^2/\hbar c \simeq 1/137$) which depend on the particular theory and are dimensionless ratios, and dimensionful `fundamental units', like the speed of light, $c$. Despite this, we can always normalise by the fundamental units to rewrite things in terms of some pure numbers. For example, normalising $c\rightarrow1$ makes a velocity, $v\rightarrow v/c$, seem like just a dimensionless ratio. However, according to at least two of the authors in \cite{Duff_2002}, the pure number $c^\prime = 1$ still contains necessary information about the structure of reality: When velocities become ${\mathcal{O}}(1)$ (in these units), we can expect dramatic new phenomena to occur~\cite{Duff_2002}. Thus, it is not quite the dimensionality that is a necessary condition, but rather the characteristic impact of the quantity on the structure and behaviour of our physical reality. In the example of $c$ this corresponds to its role as a universal limit on the velocities of particles.
Other dimensionless quantities that are not considered fundamental by our definition are mathematical constants such as $\pi$ or $e$, etc. These are ideal, abstractly defined quantities, calculable\footnote{As opposed to measurable.} to arbitrary precision, with values following from our mathematical description, that could in principle exist independently of the current instance of the universe, thus are not deemed fundamental constants.

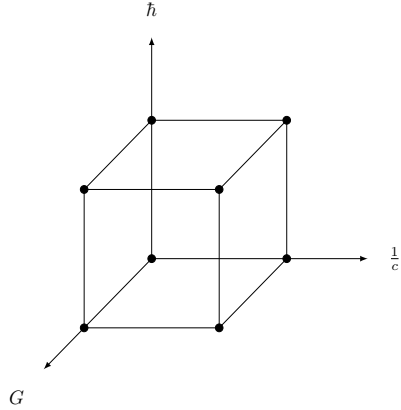
\begin{figure}
\centering
\resizebox{0.32\linewidth}{0.32\linewidth}
{
\begin{tikzpicture}
		\node (0) at (0, 0) {};
		\node  (1) at (4, 0) {};
		\node(2) at (0, 4) {};
		\node (3) at (-2, -2) {};
		\node   (4) at (0, 4.5) {{$\hbar$}};
		\node   (5) at (4.5, 0) {{$\frac{1}{c}$}};
		\node   (6) at (-2.5, -2.5) {{$G$}};
		\node   (7) at (2.5, 2.5) {};
		\node   (8) at (-1.25, 1.25) {};
		\node   (10) at (1.25, 1.25) {};
		\node   (11) at (0, 2.5) {};
		\node   (12) at (2.5, 0) {};
		\node   (13) at (-1.25, -1.25) {};
		\node   (14) at (1.25, -1.25) {};
		\draw[->]  (0.center) -- (2.center);
		\draw[->] (0.center) to (1.center);
		\draw[->] (0.center) to (3.center);
		\draw (11.center) -- (7.center);
		\draw (7.center) -- (10.center);
		\draw (10.center) -- (8.center);
		\draw (8.center) -- (11.center);
		\draw (8.center) -- (13.center);
		\draw (13.center) -- (14.center);
		\draw (14.center) -- (12.center);
		\draw (12.center) -- (7.center);
		\draw (10.center) -- (14.center);
        \filldraw (0, 0) circle (2pt) {};
        \filldraw (2.5, 2.5) circle (2pt) {};
		\filldraw (-1.25, 1.25) circle (2pt) {};
		\filldraw  (1.25, 1.25) circle (2pt) {};
		\filldraw  (0, 2.5) circle (2pt) {};
		\filldraw (2.5, 0) circle (2pt) {};
		\filldraw (-1.25, -1.25) circle (2pt) {};
		\filldraw (1.25, -1.25) circle (2pt) {};
\end{tikzpicture}
}
\caption{Graphical depiction of the cube of physical theories in the space defined by the orthogonal axes of $G$, $1/c$, and $\hbar$.}
\label{f:cube}
\end{figure}

One convenient and compact way to talk about our current framework for understanding reality was proposed by Zelmanov~\cite{zelmanov1967kosmologia,zelmanov1969}, following from works by Gamow, Ivanenko and Landau~\cite{Gamov1928}, and Bronshtein~\cite{Bronst,bronshtein1936}. For much better reviews of the subject material, see~\cite{gorelik1983,okun1991,OKUN_2002,Gorelik:2005an}. 
The system is called the `cube of physical theories' and imagines the space described by three orthogonal axes, each measuring one of the following: the inverse speed of light in a vacuum, $1/c$, the gravitational constant, $G$, and the reduced Planck's constant, $\hbar$ (see \cref{f:cube}). In this depiction, each vertex represents a different potential physical theory, not all of which had then been, or are now, fully described~\cite{Gamov1928,zelmanov1967kosmologia,zelmanov1969,gorelik1983,Gorelik:2005an,okun1991,OKUN_2002}. For example, the vertex at the origin represents classical, non-relativistic mechanics, while the vertex that lies along the $1/c$ axis corresponds to special relativity. The final goal in this picture is to develop the theory at $(1/c,\,\hbar,\,G)$, which would appear to be all-encompassing, and hope that, in so doing, the fundamental parameters be simultaneously determined as as consequence of self-consistency \cite{1983RSPTA.310..249W,okun1991}.

The question of which fundamental constants are indeed fundamental is still a matter of debate, and the consensus in definition and qualities of the fundamental constants given above is a result of some compromise on this author's part. As for the actual number, depending on who you ask, and which of the components of the definition one is willing to sacrifice, there are various arguments for variously 3, 2 or 0 fundamental constants, among others~\cite{Duff_2002}. 
Clearly, it is a complex issue that has yet to be satisfactorily resolved. Here we are not concerned so much with the absolute number of fundamental constants, because that is too difficult, but rather a lower bound on the number of fundamental constants that it is conceivable to have in our description of reality. The answer is, of course, one, as will be explained in the detailed and incontrovertible arguments set out in the next section.

\section{The Least Fundamental}

We introduce the quantity \neun\footnote{Pronounced `nee-uhn', this represents the voiced alveolar nasal consonant, similar to English `n', in the Korean hangeul script.} which describes the total number of fundamental constants required\footnote{The total number of `fundamental' instances required in writing this text, is 62.} by our physical reality in order to be described by a self-consistent mathematical framework. 
With regards to the cube of physical theories, \neun refers to the number of axes in this picture, or the dimensions of this space, required for a complete description of our physical reality. 

In order to determine a lower bound, we start with some basic statements that we believe the sensible reader will consider reasonable. First, it should be clear that \neun is an integer, since we cannot have partial constituents of a fundamental constant, by definition. Second, we are positive that \neun cannot be negative, as this does not really have any meaning. 
Then, by simple examination, we have min(\neun)$= 0$.

However, if \neun were itself a fundamental constant\footnote{As a fundamental constant, \neun would be difficult to represent directly as another axis in the scheme of the cube of physical theories, but that is not sufficient grounds for dismissal of this postulate.} (something which will be proved unambiguously to the reader in the remainder of the text) then this value would be impossible, since we would know that there is always at least one fundamental constant, \neun itself, implying that the lower bound is in fact min(\neun)$= 1$.

To begin the final task of proving that \neun is in fact fundamental, we first consider the component in the definition which relates to dimensionality. \neun is clearly a pure number, without apparent dimensions. However, upon closer inspection, we see that it behaves in a similar way to the other fundamental constants. Imagine, for a moment,  the case \neun$= 4$, accounting for the set \{$c$, $G$, $\hbar$, \neun\unskip\}. We can rescale this quantity by the value of \neun, so that \neun $\rightarrow$ \neun\unskip${}^\prime = 1$. Then, if we consider the special case where we are constrained to one of the axes of the cube of physical theories (i.e. there is only \neun\unskip${}^\prime$ and one other fundamental parameter) we would be in a regime where the rescaled number of fundamental constants take a value of $1/2$. It is only once that number approaches ${\mathcal O}(1)$ that the dramatic phenomena, on the vertex which requires all three of the fundamental constants in this example, can be fully realised. In this sense, \neun is not really dimensionless; rather, it has dimensions of dimension, and it is this characteristic that informs its impact on our physical reality. As a coincidental windfall, this fits neatly into the logic of the (hyper)cube of physical theories where, though there is no additional axis for it, \neun characterises the required number of dimensions of the space for the description be complete. 

The astute reader notices that, if we have the set of fundamental constants, we can simply count them, thus deriving the number of fundamental constants, which is then seemingly not fundamental. However, the even more astute reader would counter that this is circular reasoning since, in order to count the total number, we would need to know that the set of fundamental constants we started with was in fact complete. In other words, we need to start by knowing that we have all \neun fundamental constants, before counting them to trivially return \neun.

One might at this point be tempted to argue that it is not only the number of fundamental constants that is a necessary fundamental constant, but that a lower bound on this number is itself fundamental too. 
However, recall our argument for excluding mathematical constants $\pi$ or $e$. While the information encoded in \neun about the number of fundamental quantities required by our physical universe is something that needs to be measured for our specific instance of reality, the lower bound itself is more akin to a mathematical constant that depends on the internal logical structure of our description. To rephrase, the value of the lower bound encodes a fundamental characteristic of the logical framework inside which the fundamental constant \neun counts the number of fundamental constants. 

If we summarise the argument so far, the simplest statement of it is: if at least \neun is not a fundamental constant, we arrive at a paradox. Say we have a set of $n$ constants that are used in our physical theory: If this set is to be called the set of fundamental constants, we must know that it is complete, containing all the necessary elements. That is, the information about size of the set of fundamental constants must be included. If it is not already among the $n$ constants, then we include it as the quantity \neun with value $n+1$. However, this value is simply the size of the resulting set, and thus not fundamental, since it can be derived from the $n$ original constants. Thus we should remove \neun from the set of fundamental constants, and are left with a set of $n$ constants. The trouble is that then we could no longer say with certainty whether it contains all the fundamental constants, and we are back where we started.

The key to this paradox lies in the phrase `If it is not already among the $n$ constants'. This is precisely the prior knowledge that is necessary to fuel the loop that implies variously that \neun is/is not fundamental. But, if we are able to measure the number of fundamental constants without recourse to assuming that our set is complete, then we resolve the paradox and \neun fits neatly into place among the fundamental constants, with the corresponding value we have measured. 

How then can we hope to make a measurement of it? As one idea, the G\"odelian answer may be that we would need to introduce a meta-\neun that tells us the value of \neun. In this case we would simply need to measure meta-\neun. A simple inductive argument would lead once again down a rather long rabbit hole of $i+1$ parameters $({\rm meta\mbox{-}})^i$\neun, that we have neither the time nor the tortoises to explore here. 
Though it leaves our paper somewhat `incomplete', the question of the fundamental unknowability of \neun is left for future study \cite{Godel:1931fvw}. However, with each successive layer $i$, the outermost quantity $({\rm meta\mbox{-}})^i$\neun would become the new fundamental constant that sources the information, derived from it in the lower layers, meaning that there would still only ever be one \textit{fundamental} constant required for this purpose. Thus, as far as the lower bound is concerned, the argument leading to min(\neun)$= 1$ still holds.

One final question is what the value of \neun, or indeed min(\neun) tells us about the nature of our reality. Another way to put this is, what characteristics of our Universe allow us to make the above arguments and arrive at our conclusion that min(\neun)$= 1$, and what would it look like if that were not the case? It is not immediately obvious, and for that reason is beyond the scope of the current work. However, given the intricacy of the completely logical argument required to reach this point, we can infer that it would probably look very different, with either a completely different logical structure, none that our current logic could make sense of, or indeed none at all.

\section{Inconclusion}

We have, in this brief treatise, described for the first time\footnote{Some might argue that it is always easiest to put a lower bound on something for the first time. To this, the response is that it was not done until now, so one may infer the level of difficulty from that fact.} the lower bound on the number of fundamental constants. Fundamental constants are, as the name suggests, fundamental to our description of reality, and the exact number of them is still a matter of intense debate. Whether or not it will one day be possible to convincingly measure the actual number of fundamental constants that is/are required to completely describe our Universe, through some self-consistent mathematical and physical framework, is currently not known. However, we can confidently conclude that the current lower bound on that number is 

\section*{Acknowledgements}
I am grateful to D.D. Sabatta, W. Sohn and J.A. Harvey for invaluable feedback on the various drafts of this work, ensuring that, though incomplete, it contains at least some truth.

\printbibliography

\end{document}